\documentclass[sigconf]{acmart}

\usepackage{xspace}
\usepackage{algorithmic}
\usepackage{algorithm}

\AtBeginDocument{%
  }

\begin{document}

\title{GenTT: Generate Vectorized Codes for General Tensor Permutation}

\author{Yaojian Chen}
\authornote{Both authors contributed equally to this research.}
\email{yj-chen21@mails.tsinghua.edu.cn}
\affiliation{%
  \institution{Tsinghua University}
  \city{Haidian Qu}
  \state{Beijing Shi}
  \country{China}
}

\author{Tianyu Ma}
\authornotemark[1]
\affiliation{%
  \institution{Tsinghua University}
  \city{Haidian Qu}
  \state{Beijing Shi}
  \country{China}
}

\author{An Yang}
\affiliation{%
  \institution{Tsinghua University}
  \city{Haidian Qu}
  \state{Beijing Shi}
  \country{China}
}

\author{Lin Gan}
\email{lingan@mails.tsinghua.edu.cn}
\affiliation{%
  \institution{Tsinghua University}
  \city{Haidian Qu}
  \state{Beijing Shi}
  \country{China}}

\author{Wenlai Zhao}
\affiliation{%
  \institution{Tsinghua University}
  \city{Haidian Qu}
  \state{Beijing Shi}
  \country{China}
}

\author{Guangwen Yang}
\affiliation{%
  \institution{Tsinghua University}
  \city{Haidian Qu}
  \state{Beijing Shi}
  \country{China}
}

\renewcommand{\shortauthors}{Chen et al.}
\newcommand{\system}{\emph{GenTT}\xspace}

\begin{abstract}
Tensor permutation is a fundamental operation widely applied in AI, tensor networks, and related fields. However, it is extremely complex, and different shapes and permutation maps can make a huge difference. SIMD permutation began to be studied in 2006, but the best method at that time was to split complex permutations into multiple simple permutations to do SIMD, which might increase the complexity for very complex permutations. Subsequently, as tensor contraction gained significant attention, researchers explored structured permutations associated with tensor contraction. Progress on general permutations has been limited, and with increasing SIMD bit widths, achieving efficient performance for these permutations has become increasingly challenging. We propose a SIMD permutation toolkit, \system, that generates optimized permutation code for arbitrary instruction sets, bit widths, tensor shapes, and permutation patterns, while maintaining low complexity. In our experiments, \system is able to achieve up to $38\times$ speedup for special cases and $5\times$ for general gases compared to Numpy.

\end{abstract}



\maketitle

\section{Introduction}
Tensor permutation, the process of rearranging the indices of a multi-dimensional array, is a fundamental operation prevalent in numerous computational domains, including neural networks, tensor networks, and the Fast Fourier Transform (FFT). In neural networks, tensor permutations are essential for data layout transformations, especially in convolutional layers. In tensor networks, they facilitate the alignment of tensor indices for efficient contraction and decomposition. Despite its widespread application, tensor permutation remains challenging due to its inherent difficulty in data locality, which is heavily influenced by the tensor’s shape and the specific permutation mapping. Tensor dimensions vary widely, from regular, power-of-two shapes to irregular, high-dimensional configurations. This variability, combined with diverse permutation patterns, hinders efficient implementations. Previous studies have employed various methods to accelerate tensor permutation, including tiling \cite{wei2014autotuning} to reduce cache misses, in-place methods \cite{catanzaro2014decomposition, wu2025eithot, cheng2025ittpd} to conserve memory, GPU parallelization \cite{hynninen2017cutt, vedurada2018ttlg}, and compilers \cite{springer2017ttc, springer2017hptt} for auto-tuning. However, these methods are typically limited to regular tensors and specific permutation patterns, and the potential of instruction-level parallelism remains underutilized..

The advent of Single Instruction Multiple Data (SIMD) architectures has provided a promising solution for accelerating tensor permutations by facilitating parallel operations on multiple data elements with a single instruction. SIMD instructions contribute significantly to the peak FLOPS of modern CPUs. Failure to fully utilize SIMD results in substantial performance degradation. For well-aligned, regular permutations, SIMD offers a nearly linear speedup, reducing the complexity to $O(N/w)$, where $w$ is the length (number of elements) of a vector register. However, SIMD complicates achieving high utilization for general permutations involving long strides \cite{nuzman2006auto, anderson2015automatic}, misalignment \cite{eichenberger2004vectorization}, and residual processing. When the permutation involves non-adjacent dimensions, particularly when the last dimension with stride-1 is changed, long strides will lead to discontinuous memory accesses. Misalignment occurs when tensor dimensions do not align with the SIMD vector width, such as a dimension of size 7 or 9 with a 512-bit SIMD vector that accommodates 16 single-precision floats. This leads to residual processing, necessitating additional instructions like masked loads or stores, which increase overhead. The diversity of instruction sets, bit widths, and data types further complicates SIMD utilization, posing challenges for designing general-purpose methods. The evolution of SIMD hardware, with bit widths expanding from 128 bits (e.g., ARM NEON) to 512 bits (e.g., x86 AVX-512) and beyond, has exacerbated these challenges. Specifically, data alignment and vector utilization become increasingly difficult, particularly for complex permutations.

Previous approaches have proposed several methods for vectorized tensor permutation. Some works\cite{ren2006optimizing, franchetti2008generating, huang2010permutation, kong2013polyhedral, springer2017hptt} employed permutation decomposition to deal with long-stride permutations. A complex permutation is divided into $k$ steps which are easier to vectorize. These methods incur $k$ times the computational overhead and may not achieve performance improvement if $k > w$, length of the vector registers. Furthermore, these methods lack a standard approach to analyze or predict the decomposition outcome, since they rely on heuristic algorithms lacking optimality guarantees. Other works \cite{wei2014autotuning, lyakh2015efficient, rodrigues2018simdization} focus on specific types of permutations tailored to particular applications. Such permutations are easier to handle and offer better performance, but at the expense of generality. In conclusion, for general tensors and arbitrary permutations, there is no deterministic, efficient vectorized solution.

In this work, we introduce a novel SIMD permutation toolkit, \system, designed to generate high-performance permutation code for arbitrary instruction sets, bit widths, tensor shapes, data types, and permutation mappings. Unlike prior methods without complexity guarantee, the time complexity of our method is capped at $O(\frac{Nlog_2w}{w})$ in the worst cases, providing deterministic $\geq w/log_2w$ times complexity reduction. Furthermore, \system fully leverages SIMD parallelism, and employed pipeline-level overlapping between memory access and register manipulation, which enables it to obtain a higher speedup than the theoretical ratio. By dynamically adapting to the target architecture and permutation requirements, our toolkit ensures efficient vector utilization without requiring multiple passes over the data or restrictive assumptions about tensor regularity. Compared to HPTT \cite{springer2017hptt}, NumPy, and PyTorch, our method achieves up to $38\times$ speedup for special cases and $5\times$ for general gases. This advancement not only bridges the gap between specialized and general permutation strategies but also paves the way for scalable, high-performance tensor operations across diverse computational workloads.

\section{Background}
\subsection{Tensor Permutation and SIMD}

Tensor permutation is a fundamental operation in multi-dimensional array manipulation, involving the reordering of a tensor’s indices according to a specified mapping. Mathematically, for a tensor \( T \) of shape \( (d_{n-1},\ldots, d_{1}, d_0) \) with \( n \) dimensions, a permutation is defined by a bijective function \( \sigma: \{0, 1, \ldots, n-1\} \to \{0, 1, \ldots, n-1\} \), transforming the original index tuple \( (i_{n-1}, \ldots, i_1, i_{0}) \) to \( (i_{\sigma_{n-1}}, \ldots, i_{\sigma_1}, i_{\sigma_{0}}) \). Specifically, \((\sigma_{n-1}, \dots, \sigma_1, \sigma_{0})\) is defined as the \textbf{permutation map}. In NumPy, PyTorch and some previous works, another definition is adopted, where the index tuple is \( (i_{0}, i_1,\dots, i_{n-1}) \) and the permutation map is \((\sigma_{0}, \sigma_1,\dots, \sigma_{n-1})\). These two notation is mathematically equivalent, and can be converted to each other. For notation convenience, we adopt the first version. The total number of elements, \( N = d_0 \times d_1 \times \cdots \times d_{n-1} \), remains unchanged, but the memory layout and access patterns are altered, often requiring a complete traversal of the tensor’s elements. In practice, this operation can be implemented either in-place, by swapping elements within the same memory buffer, or out-of-place, by copying elements to a new buffer in the desired order. The computational complexity of a naive permutation is \( O(N) \), as each element must be accessed and relocated once, but the efficiency heavily depends on memory access patterns and hardware capabilities. A naive out-of-place permutation can be implemented as Algorithm~\ref{alg1}, where all tensors are organized as long 1D arrays in memory. The bijection generation is a preprocessing step that needs only to be done once. 

\begin{figure}[t]
\centerline{\includegraphics[width=0.5\textwidth]{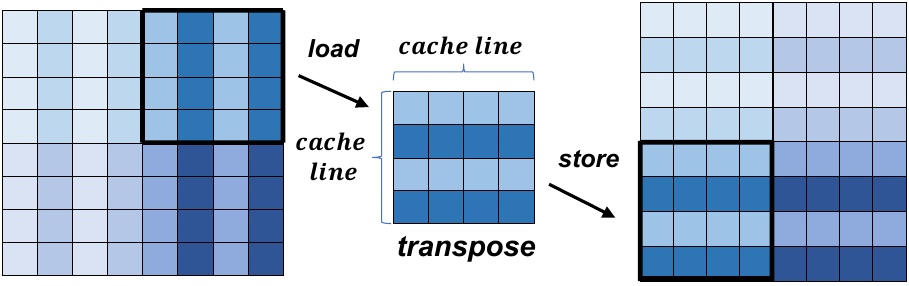}}
\caption{Matrix transposition with tiling. The whole matrix is divided into small blocks whose shape matches the size of cache line. Vectorized transposition by shuffle instructions is done on each block.}
\label{tiling}
\end{figure}

\begin{algorithm}
	\caption{Naive Tensor Permutation}
	\label{alg1}
	\renewcommand{\algorithmicrequire}{\textbf{Input:}}
	\renewcommand{\algorithmicensure}{\textbf{Output:}}
	\begin{algorithmic}
		\REQUIRE Tensor $A$ of shape $(d_{n-1}, \dots, d_1, d_{0})$, permutation map $P=(\sigma_{n-1},\dots, \sigma_1, \sigma_{0})$
		\STATE \texttt{/* Generate Bijection for All Elements */}
            \STATE $f = calc\_bijection(P)$
            \STATE \texttt{/* Relocate Elements */}
            \FOR{$i$ in $range(d_0\times d_1\times\dots \times d_{n-1})$}
                \STATE $\mathbf{B[i] = A[f[i]]}$ (or $\mathbf{B[f^{-1}[i]]= A[i]}$)
            \ENDFOR
		\ENSURE The reordered tensor $\mathbf{B}$ with shape $(d_{\sigma_{n-1}}, \dots d_{\sigma_1}, d_{\sigma_{0}})$
	\end{algorithmic}
\end{algorithm}

Although theoretical complexity of permutation is deterministic $O(N)$, the shape of tensors and the types of permutation have a significant impact on the actual performance on modern architectures. From the perspective of locality, the temporal locality of a single permutation is poor (every element will only be accessed once), while the spatial locality depends strongly on the type of permutation. For a tensor, the stride of a dimension is defined as the distance in memory between two elements that differ by 1 in this dimension. For instance, the tensor with shape $(3, 5, 7)$ has strides of $(35, 7, 1)$. After permutation, the strides change accordingly. Consequently, adjacent elements in the original tensor may be far apart in the permuted tensor. That means, according to Algorithm~\ref{alg1}, either the input $A$ or the output $B$ will not be accessed continuously.

For matrix transposition with large dimension sizes, tiling and vectorization is well-studied. The matrix can be simply tiled to be a block whose shape matches the size of cache lines. As Figure~\ref{tiling} shows, after tiling, the transposition in a small block becomes regular and cache-friendly. The blocks are designed to be fully loaded into the cache. Moreover, if the block can be completely stored in vector registers, the non-temporal instructions can be used to saving bandwidth and avoid cache pollution\cite{wei2014autotuning}. Particularly, if using $m=1$, the whole block can be loaded into $w$ registers, with inter-register data exchange by shuffling.

The irregularity of tensors with many small dimensions causes the above method to fail. The reasons are two-fold. (1) Irregular tensor shapes, such as \( (9, 3, 6, 5, 7) \), where dimensions are neither powers of two nor multiples of SIMD lengths (e.g., 4, 8, or 16), disrupt memory alignment and vector utilization. For instance, when loading a vector of 8 elements from a dimension of size 3, there will be 5 slots unused, with only $37.5\%$ efficiency. For dimension of size 9, it will be divided into $8+1$, leaving residual elements that require separate scalar processing, leading to double time cost. In addition, these unaligned memory access will overwrite the following data when storing, which strictly limits the transposition order. From cycle-based method of in-place transposition\cite{sung2014place, gustavson2019algorithms}, it is known that strict order will do harm to cache hitting and reduce performance. (2) Complex permutation maps, such as \( (1, 2, 3, 4, 5) \to (5, 3, 1, 2, 4) \), introduce non-contiguous memory access patterns with strides that vary across dimensions (e.g., from \( (630, 210, 35, 7, 1) \) to \( (810, 162, 27, 9, 1) \) in the example above). The large change in stride makes it difficult to extract blocks from large tensors. Thus, vectorization happens only on some relatively continuous pieces, which highly rely on manual specification. These irregular strides defy the contiguous or small-stride assumptions of traditional SIMD optimizations, leading to frequent cache misses and underutilized vector lanes. The lack of a unified framework to adaptively handle arbitrary shapes and mappings has left general-purpose permutation performance lagging, particularly as SIMD bit widths widen, amplifying the penalty. 

Tensor shapes and permutation types differ significantly across application domains, reflecting their diverse requirements. In artificial intelligence (AI), permutations are frequently used to reformat tensor layouts for deep learning frameworks, such as transposing a rank-4 tensor from \( (N, H, W, C) \) (batch, height, width, channels) to \( (N, C, H, W) \) to optimize convolution operations. Since the permutation map is fixed, small $C$ is the main challenge\cite{zheng2018optimizing}. In tensor networks, used in quantum physics\cite{li2021sw_qsim, chen2023lifetime}, permutations align indices for contraction, often requiring partial transpositions that may span non-adjacent axes, introducing irregular strides or permutation decompositions\cite{solomonik2013cyclops}. In Signal processing, such as in multi-dimensional FFT\cite{franchetti2007simd, xu2011high}, relies heavily on bit-reversal permutations—small-stride, requiring rule-based manually designed shuffling. In these applications, the ways of transposing tensors are complex and diverse, and are not limited to a simple pattern. Many of these transpositions involve the complex situations mentioned above. Moreover, in some scientific computing scenario like computational fluid dynamic (CFD)\cite{ferziger2002computational} and molecular dynamics (MD), there will be more irregular permutations. These difficulties underscore the need for a versatile, complexity-preserving approach to tensor permutation across diverse computational contexts.

\subsection{Related Work}
Early efforts\cite{ren2006optimizing, franchetti2008generating} studied SIMD vector permutation. As a special kind of vector permutation (if flattening tensors as 1D), in tensor permutation we can directly inherit vector-based methods. These methods introduced the concept of decomposing complicated permutations into a sequence of simpler operations, which could be easier mapped to SIMD instructions like shuffle. While effective for certain cases, this approach often increases computational complexity for highly intricate permutations, as the number of decomposition steps grows, leading to additional memory accesses and instruction overhead. In some bad situations, $k$ decomposition steps lead to a complexity of $O(kN/w)$. The total complexity may even increase after decomposition and SIMD optimization. In our observation, vector permutation and tensor permutation should be divided into two different issues, since vector-based method is less efficient for tensor permutation. There are two main reasons: (1) Vector permutation itself may not have clear rules. In contrast, tensor transposition is essentially the permutation of dimensional vectors. After being mapped to tensor memory, it is subject to strict constraints. (2) Vector-based method are not good at handling long stride problems because they involve data exchange across long-distance registers, which increases complexity. In tensor permutation, we can overcome long stride by very few additional instructions. 

Some subsequent research shifted focus toward specific permutation patterns, particularly those associated with FFT\cite{franchetti2007simd, xu2011high} and tensor contraction\cite{shi2016tensor, springer2018design, chen2025swtncreachingcomplex}, a key operation in numerical libraries and machine learning frameworks. Some works\cite{stock2011model, rodrigues2018simdization} focus on small, structured tensors, which are micro kernels of tiled tensors. \cite{lyakh2015efficient} is designed for contractions in quantum many-body problems. Permutations in tensor contractions represent a important subcategory that some indices are moved to the front or the end while keep the others unchanged. These permutations show significant complexity with long strides and misalignment. Though these advancements have been confined to predictable permutation modes, they can provide a lot of inspiration for the design of general cases.

Automation is a long-standing topic in the field of vectorization\cite{nuzman2006auto, trifunovic2009polyhedral, baghsorkhi2016flexvec, moghaddasi2024vectron}. Auto-vectorization for tensor permutation has attracted the attention of many researchers. Code generators and compilers are two primary solutions. The code generators\cite{wei2014autotuning} are designed for a series of custom permutations using manually deduced rules. The limitation of code generators is their generalizability, since they often belong to specific instruction sets and tasks. Compilers\cite{springer2017ttc} are commonly general, but lack of rule-based simplification. As research deepens, it is necessary to combine the characteristics of the two to achieve the best results\cite{kong2013polyhedral}, so the boundaries are gradually blurred. Recently, AI-driven approaches, including large language model (LLM)-based vectorization\cite{zheng2025vectrans, taneja2025llm}, show promise for regular loops but struggle with the semantic complexity of arbitrary tensor permutations.

\section{Tiling Framework}

We have discussed the traditional tiling method in matrix transposition. However, it can not be simply transplanted to tensor permutation due to its requirements in shape. When using vectorized loading and storing, the parallelly managed elements should be adjacent in memory. According to Figure~\ref{tiling}, large matrices have long enough rows before and after the transposition to ensure the spatial locality required by SIMD. In tensors, however, there are commonly some dimensions whose size is extremely small, like 2 or 3, which is far from the required continuous memory to fill a vector register. Moreover, unlike matrix transposition, which is completely fixed in form, the way tensors are permuted is highly arbitrary and can become very complex. With these complexities added, the design of tiling for tensor transposition becomes much more difficult.

Although direct tiling will fail in tensor permutation, some of its advantages and properties could be borrowed. The key success of tiling in matrix transposition comes from a small block whose elements have good spatial locality in both matrices before and after transposition. From the perspective of tensor permutation, to keep similar spatial locality, we also need to find such a block whose elements are continuous in both tensors before and after permutation. Then, the whole permutation is decomposed into a block version consisting of a high-level permutation to rearrange blocks and a low-level permutation inside each block. These blocks meet the requirement of vectorized loading and storing simultaneously. Thus, SIMD can be applied.

\begin{figure*}[t]
\centerline{\includegraphics[width=1.0\textwidth]{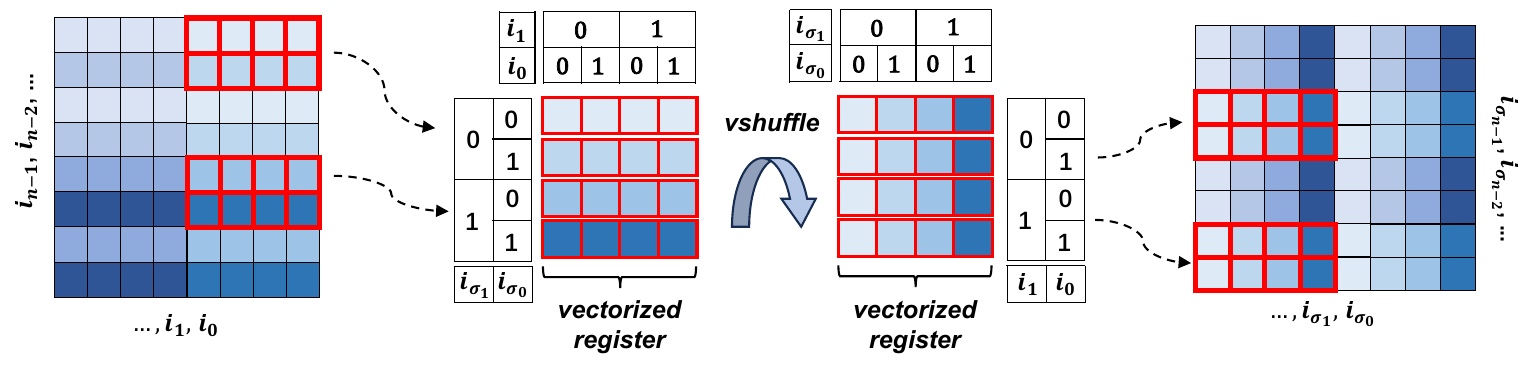}}
\caption{Tiling framework of tensor permutation, targeting tensors with all dimensions of size 2. Vector length $w=4$. The last $log_2w=2$ indices of the original tensor (i.e. $i_1, i_0$) and the permuted tensor (i.e. $i_{\sigma_1}, i_{\sigma_0}$) are combined together to form a squared block. Local permutation inner-block is implemented by vectorized shuffle. The registers can continuously read data from the memory and continuously write it back after the local permutation is completed.}
\label{caching}
\end{figure*}

To find such a block, the memory continuity is a critical target. In arbitrary tensors, only in the last dimension, the elements' proximity in coordinates is equivalent to their proximity in memory. Thus, the rows of the blocks can be formed by the last few dimensions of the original tensor, i.e. $i_0, i_1, \dots$. Symmetrically, the columns of the blocks are flattened by $i_{\sigma_0}, i_{\sigma_1}, \dots$. The shape of the blocks is designed to be similar to the number of SIMD elements. That means a proper $p$ and $q$ with $d_0\times d_1\times\dots\times d_p \sim w$ and $d_{\sigma_1}\times\dots\times d_{\sigma_q} \sim w$. This kind of block is also very cache-friendly due to its spatial locality. From here, the basic tensor permutation framework seems to be ready. However, there are still many problems like alignment, residual processing and local permutation design, preventing this method from being transformed from a theoretical framework into a practical algorithm. For example, if we get a block with shape $(6, 3, 5, 7)$ and want to do local permutation to reorganize it as $(7, 5, 3,6)$ by 8-element SIMD instructions, we will find it hard with the framework above. Although the continuity is satisfied, simply dividing the $18 \times 35$ sub-matrix as $8\times8$ tiles does not work, since there are more granular permutation $(5,7)\rightarrow(7, 5)$ and $(6,3)\rightarrow(3,6)$. In the following sections, we will divide and conquer these issues from special to general, and at the end propose a solution for arbitrary SIMD tensor permutation.

\section{SIMD Permutation for Tensors with All Dimensions of Size 2}
\subsection{Block Address Manipulation}
We first consider a special family of tensors with all dim-size $d_i = 2$. These tensors have regular shape and exactly meet the $2^n$ number of elements in vector registers. In addition, this family of tensors is widely used in FFT and quantum computing. The optimization for these tensors can be directly used to accelerate certain applications.

For tensors with all dim-size $d_i=2$, each length-$w$ vector register can store a rank-$log_2w$ sub-tensor. According to the discussion above, we need to find a $w\times w$ block with $i_0, i_1, \dots, i_p$ and $i_{\sigma_0}, i_{\sigma_1}, \dots i_{\sigma_q}$ for tiling. Since $d_i=2$, there will be $p=q=log_2w$. Thus, we need only to find the last $log_2w$ indices from the original tensor and the permuted tensor, respectively, and reshape the sub-tensor into a square. Each block holds a complete sub-tensor, so the local permutation can be done by vectorized instructions like vshuffle. The main problem is how to calculate the offset, which decides the address of each loading and storing. The offset of each vector register is just the address of its first element. In a tensor, the offset relative to element 0 in the memory of an element with coordinate point $(a_0, a_1, \dots, a_{n-1})$ is:
\begin{equation}
    offset=\sum_{k=0}^{n-1}\limits a_k2^k, a_k \in \{0,1\}
\end{equation}
After permutation, its new coordinates will be transformed to $(a_{\sigma_0}, a_{\sigma_1}, \dots, a_{\sigma_{n-1}})$. The address of Figure~\ref{caching} illustrates an example with $w = 4$. The block consists of 4 indices $i_0, i_1, i_{\sigma_0}, i_{\sigma_1}$.  According to the tiling scheme, the high-level permutation is done block-wise. As a result, the address of each register of a certain block can be divided into the block address and the register offset: $reg\_add = block\_add + reg\_off$. The block address is the beginning address of a block, i.e. the address of the first element of the block. Such address can be determined by taking the 0-th component of all intra-block indices, i.e. $(i_0, i_1, i_{\sigma_0}, i_{\sigma_1})$. Extending to the general $w$, letting the set of intra-block indices be $I_B$, the block addresses in the original tensor and the permuted tensor are:

\begin{gather}
    block\_add_{ori} = \sum_{k\notin I_B}^{n-1}\limits a_k2^k, a_k \in \{0,1\} \\
    block\_add_{perm} = \sum_{\sigma_k\notin I_B}^{n-1}\limits a_{\sigma_k}2^k, a_{\sigma_k} \in \{0,1\}
\end{gather}

In a block, registers are numbered according to column indices. Before permutation, indices $i_{\sigma_0}, i_{\sigma_1}, \dots, i_{\sigma_{log_2w}}$ encode the order of vector register. For example, in the left half of Figure~\ref{caching}, each of $00, 01, 10, 11$ of $(i_{\sigma_0}, i_{\sigma_1})$ refers to a register. Symmetrically, the register offset after permutation is decided by $i_0$ and $i_1$. For general $w$, the register offset can be written as:
\begin{gather}
    reg\_off_{ori} = \sum_{k\in\{i_{\sigma_0}, \dots, i_{\sigma_{log_2w}}\}} a_k2^k, a_k \in \{0, 1\} \\
    reg\_off_{perm} = \sum_{k\in\{i_{0}, \dots, i_{{log_2w}}\}} a_{\sigma_k}2^k, a_{\sigma_k} \in \{0, 1\}
\end{gather}

It is worth noticing that the index calculation is actually very time-consuming even in the context of scaler tensor permutation, since it requires element-wise operations. For each element, to traverse the value of each dimension takes $O(logN)$ time. This overhead will be even larger than data movement. If the permutation needs to be executed repeatedly for tensors with same structure, one can use pre-calculated lookup table, sacrificing some memory. Here we use a recursive method. We classify all blocks by the number of $1$ in its binary block address. Starting from the only block in the 0-th class, we only need to do one more calculation $a_k2^k$ to get the $k$-th block in the 1-st class. Recursively, addresses of all blocks could be calculated in $O(1)$ time, and so as the register addresses. 

\subsection{Local Permutation by Shuffle}
\begin{figure}[t]
\centerline{\includegraphics[width=0.5\textwidth]{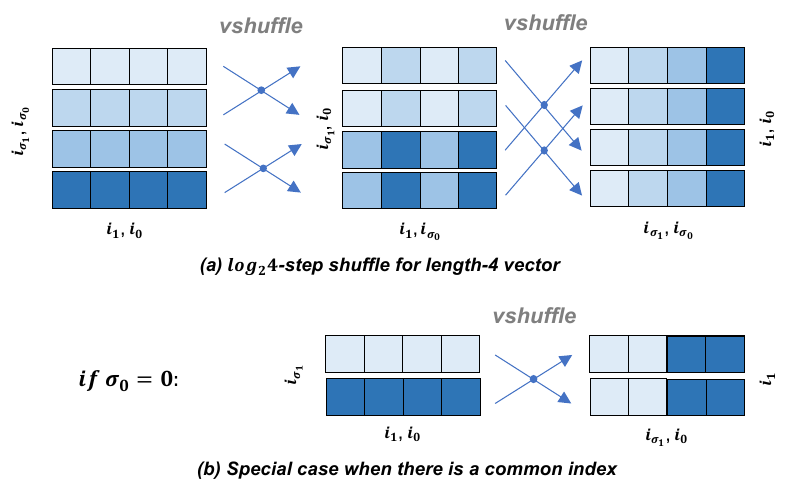}}
\caption{Local permutation by multi-step shuffle. $w=4$. (a) The worst case with $log_24 = 2$ steps. 4 registers are used. At step 1, $i_0$ and $i_{\sigma_0}$ are exchanged. At step 2, $i_1$ and $i_{\sigma_1}$ are exchanged. (b) A special case when $\sigma_0 = 0$ with only one step.}
\label{shuffle}
\end{figure}

Tiling, loading and storing are sufficiently discussed above. In this section we propose the local permutation inside each block. Figure~\ref{shuffle} (a) provides an example using length-4 vectors. After loading, a block with 4 indices $i_{\sigma_1}, i_{\sigma_0}, i_1, i_0$ is held by 4 vector registers. Initially, in each register there is a sub-tensor with indices $i_1, i_0$. After two steps of vectorized shuffle, the block is permuted with new indices order $i_1, i_0, i_{\sigma_1}, i_{\sigma_0}$.

Vectorized shuffle (vshuffle, vshuf, vperm) is a widely used instruction family in modern architectures. These instructions allow element selection from one or two source vector registers, controlled by an index vector. The typical usage of vshuf is:  
\begin{verbatim}
                 vshuf a, b, c, d
\end{verbatim}
in which $a$ and $b$ are source vector registers, $d$ is the target vector register. $c$ is the shuffle index, indicating which position of the source vector each element of the target vector comes from. 

Comparing the initial data layout and the target data layout in each register in Figure~\ref{shuffle} (a), we can find that finally all registers collect one element from the others. Recalling that the shuffle instructions deal with two-vector element reordering, the collective operation should be implemented by multiple steps. These data exchange between registers are very similar with AlltoAll communication. Here we borrow the butterfly network method\cite{avior1996tight, izzi2023realizing} to do this communication. There are $log_2w$ communication steps in total for $w$ registers. In the $k$-th step, the $i$-th register will be shuffled with the $i\oplus 2^k$-th register, where $\oplus$ denotes to element-wise XOR. From the perspective of tensor permutation, in each step, the shuffle instruction can swap two indices and do inner-register permutation simultaneously. Taking a comprehensive view, a column index swaps the position with a row index. Then, the indices in a register are optionally reordered. These two permutations are composited together to be one shuffle.

In the shuffled block, the new row indices are properly reordered, but the new column indices may not be correctly sorted. Here we introduce the register-rename method. Although the vector registers are organized as a block, the order of them is flexible. For example, if the order of $i_{0}$ and $i_{1}$ should be swapped after permutation, we can simply rename the registers to exchange them implicitly. That means the register number will be adjusted from $00, 01, 10, 11$ to $00, 10, 01, 11$. 

When there are common indices between the row indices and the column indices, there will be some simplification. As Figure~\ref{shuffle} (b) shows, when $\sigma_0 = 0$, there is one common index, then the block becomes a rank-3 tensor. Since only two vector registers are needed, the communication can be done with in one step. In general, each common index reduces one shuffle step. In particular, if the row indices are exactly same as the column indices, there is no shuffle instructions needed.

\subsection{Shuffle Index Generation}
In a shuffle step, there are $w$ shuffle instructions in total. However, that does not mean we should generate $w$ shuffle index vectors. Registers are grouped to do pair-wise communication, and the behavior of every group is same. In a pair, the two shuffle operations are symmetric. That means we only need to calculate one shuffle index vector for each step, and then generate its pair by symmetry. For convenience, before the last step, we swap $i_{\sigma_k}$ with $i_{k}$. At the last step, we do a composite shuffle to swap $i_{\sigma_0}$ and $i_0$, and reorder all new row indices. Thus, the shuffle indices of the first $\log_2w -1$ steps are fixed for all kinds of permutation, and we should only deal with the last step. For the last step, the shuffle index is almost the index of a length-$w$ vector permutation. The only difference is that, each two adjacent elements come from different registers. As a result, with some built-in shuffle index vectors, for every permutation we need only calculate one shuffle index vector.

\section{SIMD Permutation for Arbitrary Tensors}

\subsection{Extension to Arbitrary Tensors}
In the previous section we have discussed the solution for tensors with all dimensions of size 2, which is a special family. In this section, the solution will be extended to arbitrary tensors. 

For tensors whose sizes of all dimensions are powers of 2, the extension can be performed by dimension decomposition. If there is a tensor $T$ with shape $(2, 16, 8, 4)$ and the corresponding indices $(i_3, i_2, i_1, i_0)$, we can reshape it to be a rank-10 tensor with all dimensions of size 2. Each index is decomposed into several sub-indices whose size equals 2. For example, $i_1$ will be sliced as $i_{12}, i_{11}, i_{10}$ with shape $8\rightarrow(2,2,2)$. Since reshape is a $O(1)$ operation with no change to memory layout, the permutations on tensor $T$ are equivalent to permutations on the reshaped tensor. Considering a permutation with map $(0,2,1,3)$, the new shape and the new indices order of $T$ will be $(4, 8, 16, 2)$. Within each dimension, the order of the elements will not change. As a result, the order of decomposed indices is also kept. Thus, the new indices order of the reshaped tensor will be $(i_{01}, i_{00}, i_{12}, i_{11}, i_{10}, \dots)$. With this decomposition, our method can be extended to tensors whose sizes of all dimensions are powers of 2 with no additional modifications required. 

For general cases, we will still discuss high-level permutations between blocks and local permutations within blocks separately. The first step is to decide which indices are chosen to form the block. In order to ensure the independence of the two levels of permutation, there should not be common indices between the high-level permutation and the local permutation. In addition, according to the index-swap shuffle scheme, indices in the block should be expanded to power of 2. The expansion of $d_i$ is done as $2^{\lceil log_2d_i\rceil}$ by ceiling. That means $2^{\lceil log_2d_0\rceil}\times 2^{\lceil log_2d_1\rceil}\times\dots\times 2^{\lceil log_2d_p\rceil} \leq w$ and $2^{\lceil log_2d_{\sigma_0}\rceil}\times 2^{\lceil log_2d_{\sigma{1}}\rceil}\times\dots\times 2^{\lceil log_2d_{\sigma_q}\rceil} \leq w$. The only exception is when the last index itself $d_0 > w$ or $d_{\sigma_0} > w$, which degenerates to matrix transposition, an already solved problem. 

Padding to power of 2 prevents us from fully utilizing the parallelism of vector registers. Some extreme cases, like $d_0=3, w=8$, will lead to only $37.5\%$ efficiency. To alleviate this problem, we applied dimension composition. Noticing that there is $2^{\lceil log_2a\rceil}2^{\lceil log_2b\rceil} \geq 2^{\lceil log_2ab\rceil}$, merging two dimensions as much as possible can reduce the performance waste of padding. If neither adjacency nor order of two indices $i_m, i_{m+1}$ is changed after permutation, they can be merged as a new index. Here are two typical examples to show the advantage of merging. For $d_0 = 3, d_1 = 5, w = 16$, it will be organized as $4\times8 = 4\times 4 \times 2$ in the original strategy, The efficiency is only $15/32$. However, if do merging, the new index will have $d=15$, improving the efficiency to be $15/16$. For $d_0 = 9, d_1 = 8, w=8$, originally $d_0$ will be divided into $9 = 8 +1$, which leads to $7/16$ performance loss. After merging, the new index $d=72$ is divisible by 8, with full performance. This method is implemented as a pre-reshape without change to the main framework.

In high-level permutation between blocks, the only change comes from the calculation of block address. In our general tensor permutation scheme, we do not need any additional data reorganization on the in-memory tensors. Since the tensors in the memory keep their layout without padding, the $2^k$s should be replaced by the strides of the tensors, and the range of $a_k$ varies from $\{0, 1\}$ to $\{0, d_k-1\}$. As for the local permutation, however, the condition becomes complex. The two main problems are unaligned memory access and padded shuffle, which will be discussed in the following sections.

\subsection{Unaligned Memory Access}
\begin{figure}[t]
\centerline{\includegraphics[width=0.5\textwidth]{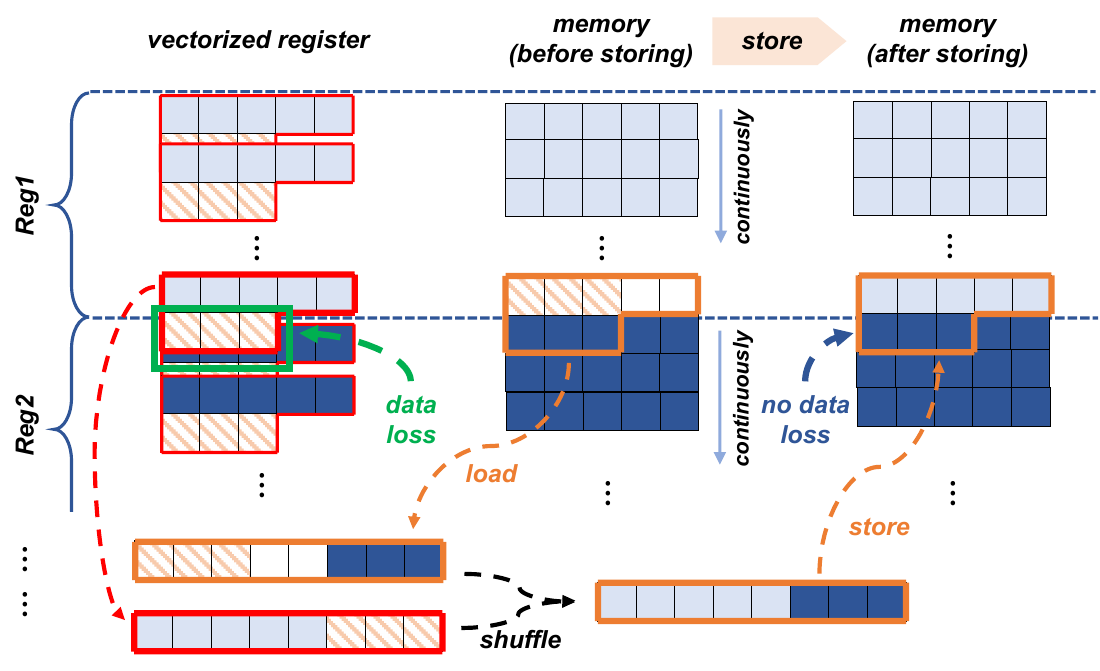}}
\caption{The unaligned storing avoiding unsafe overwriting. $w=8$. Valid data is marked in blue (dark blue and light blue refer to different registers). The others are invalid. Assuming $d_{\sigma_0}=5$. Then there are 5 valid elements in each register. \emph{Reg1} denotes the data stored by register 1. "Before/after storing" refers specifically to "before/after the out-of-bound storing". }
\label{store}
\end{figure}

In order to maintain the original memory layout, dimension expansion is only performed in blocks. When a length-8 vector register tries to load data from the tensor whose last dimension only has size of 7, there will be unaligned memory access. Thanks to the \emph{uload} and \emph{ustore} instructions without performance loss in arm and x86 CPUs, we do not need to design unaligned operations. If the register should be responsible for a $3\times7$ sub-tensor, the beginning address of all loading instructions will be $0, 7, 14$. Within each loading, there will be 7 valid elements, with the last one redundant. During loading, the actual data layout in the memory and the data organization for shuffle is different if the register should load more than 1 indices. For example, two indices with shape $2\times3$ is padding to be $2\times 4$ and loaded into a length-8 vector. The expected layout in the register is: 3 valid elements, 1 invalid element, 3 valid elements and 1 invalid element. However, in the memory the 6 valid elements are continuously arranged. An extra self-shuffle is applied to distribute the data from \{6, 2\} to \{3, 1, 3, 1\}. Almost same thing happens when storing, except for some boundary conditions.

Figure~\ref{store} demonstrates the boundary condition of overwriting. With 5 valid elements, the workspaces of one length-8 register are overlapped. In the first iteration, the register stores 5 valid elements and 3 redundant elements. The 3 redundant element is then overwritten in the next iteration. This overwriting is safe, since the valid elements cover invalid ones, as the top left corner of Figure~\ref{store} shows. However, when the process goes to the boundary of the area that two registers are responsible for, the unsafe overwriting happens. In the center left part of Figure~\ref{store}, we can find that, during the last iteration, the 3 redundant elements will be written out of bounds. The other side of the boundary is the valid data already written in the first iteration of reg 2. Thus, this overwriting will lead to data loss. According to the memory layout before storing of the last iteration, the valid data of reg 2 already exists. So, the solution is to reserve the data in another register, and then do reorganization to collect all valid data together in one register. This turns the last 3 elements to be valid, and ensures the security of out-of-bound storing.

\subsection{Padded Shuffle}
The padded shuffle scheme inherits the squared shuffle scheme illustrated in Figure~\ref{shuffle}. The main framework is unchanged, but the padding nature helps to reduce operations. After local permutation, the valid elements are distributed at the front of the registers and the first several registers. This means that the data in the last few rows and columns of the block are invalid. The operations corresponding to these halos can be canceled. Specifically, we investigate all shuffles, a shuffle will not be executed if the output vector contains no valid element. The auxiliary registers are also virtualized to save registers. In practice, for shuffle between $a$ and an auxiliary register $b$, it is treated as a self-shuffle of $a$ with same shuffle index.

\section{Automation}
\subsection{Cross-Platform Code Generator}
To ensure that tensor permutation can be executed efficiently across diverse hardware platforms, we designed a cross-platform code generation backend capable of targeting multiple instruction sets. The goal is to abstract the permutation operations and the architecture-specific details while maintaining high performance and portability.

Our code generator is structured around a modular \textbf{intermediate representation} (IR), which captures high-level tensor operations, shuffle strategies, and memory access patterns in a platform-agnostic way. This IR is then lowered into target-specific code depending on the hardware selected. It accepts parameters of machine and program to generate customized code. The current supported parameter space is shown in Table~\ref{tab:parameters}.

\begin{table}[]

    \centering
    \begin{tabular}{c|c}
       Parameters  &  Space \\
       \midrule
        Instruction Set & x86 AVX; ARM SVE; Sunway SIMD \\
        Bitwidth & 128, 256, 512 \\
        Datatype & $sizeof(type) \geq 32$\\
        Tensor shape & Arbitrary tensors\\
        Permutation map & Arbitrary maps
    \end{tabular}
    \caption{A list of parameter space supported.}
    \label{tab:parameters}
\end{table}

The key function of IR is to decouple the permutation strategy and the specific hardware. Thus, it is a hardware-independent representation.  According to the previous discussion, the permutation strategy can in fact exist independently of the specific hardware and is only related to the number of elements in the vector register. Information on data type, SIMD width and instruction set architecture (ISA) of the upper-level strategies is hidden and condensed into the vector length $w$. IR contains a series of abstract operations such as \emph{load}, \emph{store} and \emph{shuffle}. For a certain tensor shape and a permutation map, given the vector length, the code generator applies the methods introduced in the previous sections to generate a strategy, which is represented as IR operations. In detail, the IR generation workflow consists of five main steps: (1) Shuffle pattern recognition to decide the number of registers and shuffle steps. (2) Permutation division to generate the \emph{for} loop for high-level permutation between blocks. (3) Dealing with padding and unaligned processing and determine the additional operations. (4) Pre-calculating shuffle index vectors. (5) Following the pattern, collecting all operations to generate IR for local permutation.

Hardware-related parts are implemented as backends in IR operations. IR operations are highly simplified representations. A series of parameters need to be determined before they can be converted into practical code. Basically, we need ISA, data type and SIMD width. The vector length $w$ is calculated as: $w = width / sizeof(type)$. In particular, on Sunway architecture, the number of elements of vector registers of float32 is the same as float64. However, considering that permutations only consist of data movement with no calculation, the data type is flexible. It is valid to use word shuffle to deal with floating points with a simple pointer conversion. So, the shuffle instructions are uniformly based on the 32-bit word interface. The generator emits architecture-specific intrinsics to fully leverage SIMD capabilities. On x86, the IR will transcript into AVX-512 instructions like \texttt{\_mm512\_loadu\_epi32}, \texttt{\_mm512\_shuffle\_epi32}. On ARM, it targets SVE and SVE2 instructions such as \texttt{svld1}, \texttt{svtbl2}. On Sunway, it uses custom shuffle/move/load primitives. Moreover, the mode of memory access also affects the instruction selection. The backends will identify unaligned memory access by the address, and call unaligned instructions like \texttt{loadu} and \texttt{storeu}. 

The IR is designed to be extensible, allowing for the addition of new instructions as new architectures emerge. To add a new architecture or some new instructions, the only thing to do is to add the corresponding logic into the backends of IR operations. This modularity enables the generator to adapt to different instruction sets and vector widths without requiring significant changes to the core logic. 

\subsection{Instruction Reorganization}
Beyond direct code generation from the permutation strategy, the framework can do some post-processing for optimization, including register reuse, loop unrolling and instruction reordering. For long vectors with $w=16$, forming a square block requires large number of registers. Without fine-grained reuse, it takes 16 registers to carry data, 16 registers to store the shuffle results, and another 8 registers to store shuffle index if there are 4 steps in total. 40 vector registers exceed the hardware resource of most machines, making the method impractical. Considering that at each step, all registers are paired, and each pair only participates in two shuffles, the two paired registers can be free after these two shuffles. Thus, we actually need only two additional registers to store the intermediate results. Thus, the number of registers is reduced to $16 + 8 + 2 = 26$.

For blocks with very few steps, the data dependence will cause damage to pipeline parallelism. For example, in a 1-step shuffle, there are only two registers, and the operations will be \texttt{load, load, shuffle, shuffle, store, store}. Noticing that shuffle depends on loading, and storing needs the results of shuffle, there are very restricted space for pipeline. Loop unrolling provides a chance to release data dependence by introducing independent instructions from other iterations. In a single loop, loop unrolling can be decided based on the required number of registers and the number of available hardware registers. Our loop unrolling is combined with instruction reordering. All the data loading and offset calculations are placed first. Then, the shuffle operations are arranged according to the order of loading instructions. Specifically, the earlier the input vectors are loaded/generated, the earlier the shuffle is executed. This reorganization strategy improve performance by minimizing pipeline bubbles.

\begin{figure*}[t]
\centerline{\includegraphics[width=1.0\textwidth]{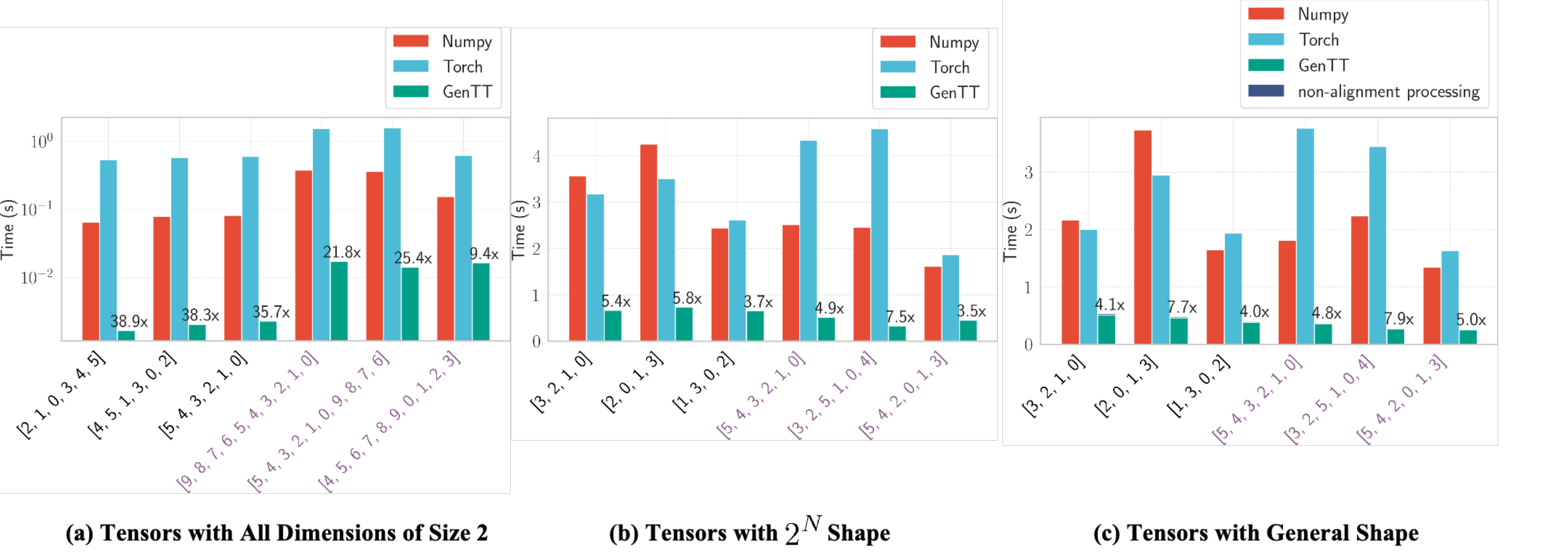}}
\caption{Time results and speedup ratios of tensors with different shapes. Two shapes with three permutation maps are selected in every sub-figure. The permutation targets are shown in the abscissa while the origin order is like [0, 1, 2, 3, 4, 5], [0, 1, 2, 3] and so on. (a) Tensors with all dimensions of size 2. The shape of the first three is [2, 2, 2, 2, 2, 2] and that of the last three is [2, 2, 2, 2, 2, 2, 2, 2, 2, 2]. (b) Tensors with $2^N$ shape. The shape of the first three is [8, 32, 32, 4] and that of the last three is [4, 4, 8, 8, 4, 4]. (c) Tensors with general shape. The shape of the first three is [7, 32, 32, 3] and that of the last three is [5, 3, 7, 8, 4, 4]. The shapes are mutated from $2^N$ shapes.}
\label{fig:shape}
\end{figure*}

\section{Evaluation}

Our experiment will be conducted on the Intel(R) Xeon(R) Gold 6230R CPU @ 2.10GHz, KunPeng 920F, and SW26010-Pro, corresponding to the x86, ARM, and Sunway platforms, respectively. Our method will be evaluated on a single core of every chip above, which is a simple and objective way to show its advantages. 

As mentioned in the previous sections, the notation we adopt about the permutation map is different from that in NumPy and PyTorch. To align the semantics, in the evaluation section we use the notation same as NumPy and PyTorch. The conversion from the previous maps to NumPy maps is: \((n-1, \dots, n-1, n-1) - (\sigma_{n-1}, \dots, \sigma_1, \sigma_{0})\). For example, $(3, 1,0, 2)$ is transformed to $(0, 2, 3, 1)$.

\subsection{Validation by Random Permutation}
To ensure the correctness of the generated code across various tensor shapes, layouts, and transformation strategies, we designed a validation pipeline focused on exact equality of results.For each generated kernel, we constructed test cases where both inputs and expected outputs were computed using \texttt{NumPy}'s native \texttt{transpose()} function. The outputs of the generated code were then compared to the NumPy reference outputs using exact bitwise equality. In all tested scenarios, the results matched perfectly, confirming the correctness of the generated indexing logic and memory access patterns.To ensure robustness across a wide spectrum of use cases, we implemented a randomized testing framework. Tensor shapes ranging from small (e.g., rank-2, rank-3) to high-dimensional tensors (up to rank-16), and permutations including identity, reversed, and mixed orders were randomly sampled. More than 1,000 test cases were generated and verified for exact equality. All tests passed without any mismatches.We further evaluated correctness across multiple architectures, including x86, ARM, and Sunway platforms. The same input tensors were used to execute the generated kernels on different hardware targets. Output tensors were then compared by element to ensure platform-independent determinism. All results were fully consistent across platforms, validating both correctness and portability.

\subsection{Performance}
\subsubsection{Tensor Shapes}

We conducted performance tests on various tensor shapes on the x86 platform. Several representative tensor shapes are selected, including tensors with all dimensions of size 2 (commonly used in quantum computing), power-of-two shape tensors (frequently used in deep learning systems) and general-shape tensors. One reason to choose these three categories is that our framework starts with the case where each dimension is 2 and gradually expands to any tensor. Another consideration is that according to our theory, when each dimension of a tensor is large, it is closer to matrix transposition, making the existing method perform better. At the same time, for any tensor, our method has performance loss caused by padding, and the performance will be relatively worse. We hope to verify these predictions through experiments.

Figure~\ref{fig:shape}(a) presents the performance results of tensors with all dimensions of size 2, whose dimensions are 6 and 10, under different permutation strategies. When the permutation map is dispersed, the normal SIMD strategy can hardly play a role because the sizes of each dimension are too small. So our method can achieve a great speedup. Figure~\ref{fig:shape}(b) shows the performance of $2^N$ shape tensors, which are common in deep learning inference workloads. Thanks to the tiling strategy, spatial locality is enhanced, allowing better use of caches across various loop structures to accelerate data access. In this scenario, an average speedup of $5\times$ is reached.

There are two possible factors that cause performance degradation in the general cases, from unaligned processing and padding, respectively. The former comes from the conditional overhead from store overwrite checks, and the latter is due to the unused slots in vector registers. When indices related to read/write operations cannot fully utilize the registers, performance discounts occur. In contrast, dimensions unrelated to read/write are batch-processed, and misalignment does not introduce additional penalties. In our experiments, we carefully test the additional overhead by unaligned operations. From Figure~\ref{fig:shape}(c), the overhead of non-alignment processing accounts for a very small portion of the total runtime, which means the performance decay mainly comes from padding. This inspires us to employ more optimizations to fill the slots of vector registers. Overall, the average speedup remains around $4\times$.

According to these tests, another observation is that, different permutation maps correspond to different performance results. In our previous discussions, the permutation maps affect performance by the number of shuffle steps in the load-shuffle-store workflow. More shuffles lead to higher per-step overhead. In some extreme tiling cases, register pressure increases, resulting in performance degradation. For general tensor shapes, the generated code may include self-shuffle overhead. The specific impact of the shuffle step will be discussed in depth in subsequent experiments.

\subsubsection{Size of Dimensions}

\begin{figure}[t]
\centerline{\includegraphics[width=0.5\textwidth]{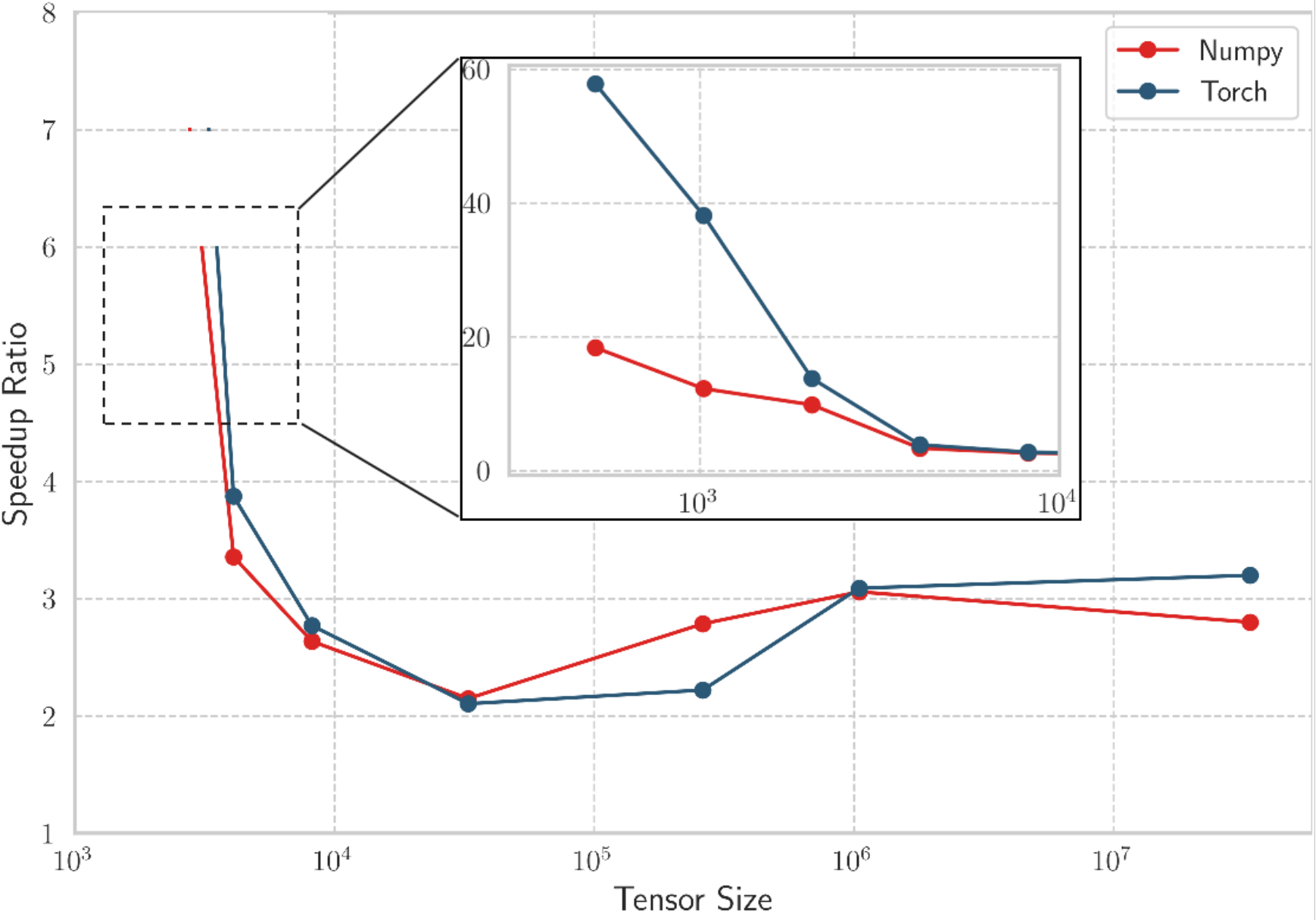}}
\caption{The speedup ratio related to the absolute size of dimensions. All tensors are rank-3 with the same size of all dimensions. When the tensor size is moderate or small ($\leq 2048$), the speedup ratio gradually decreases from $60\times$. As the tensor size increases, the speedup ratio stabilizes at around $3\times$.}
\label{fig:size}
\end{figure}

Here we will discuss the impact of dimension size on performance. To control other variables, we test rank-3 tensors with the same size of all dimensions and fixed permutation map similar with matrix transposition. This pattern refers to permutations in deep learning. Figure~\ref{fig:size} illustrates the speedup of the auto-generated code compared to \texttt{NumPy} and \texttt{PyTorch} as the absolute size of the tensor increases. When the tensor size is moderate or small (i.e., size $\leq$ 2048), the speedup can reach approximately $10\times$ in average. In spite of the SIMD optimizations in \texttt{NumPy}, even for large-scale tensors, a speedup of around $3\times$ can still be achieved. The use of shuffle-based data movement workflow is shown to be more efficient than scalar and \texttt{NumPy}-like SIMD data transfers. At the same time, the results also show that the existing work is relatively more efficient for the implementation of large-scale tensor permutations, while the optimization of small-scale tensor permutations is still insufficient.

\subsubsection{Register Utilization and Instruction Reordering}

\begin{figure}[t]
\centerline{\includegraphics[width=0.35\textwidth]{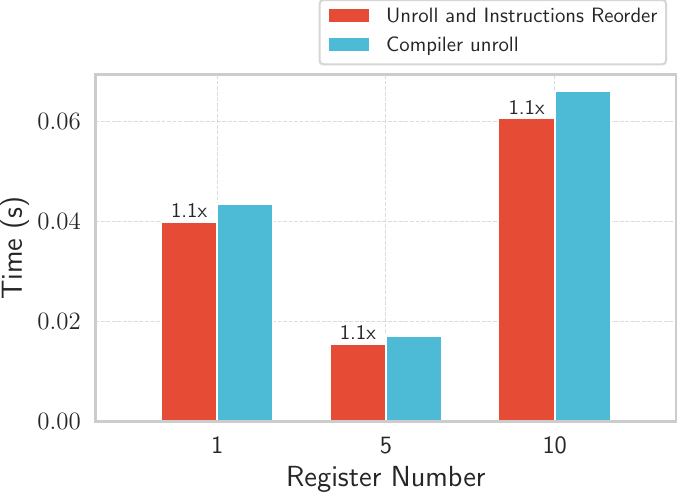}}
\caption{The performance improvement of loop unrolling and instruction reordering compared to compiler automatic unrolling. The x-axis represents the number of vector registers used in a single loop.}
\label{fig:unroll}
\end{figure}

Under certain permutation map, a few vector registers are required within a single loop. For example, permutation configurations such as permutation map \texttt{(2, 1, 0, 3)} and tensor shape \texttt{(64, 32, 32, 4)} require only 5 vector registers. Manual loop unrolling and instruction reordering are needed to utilize additional vector registers then fully exploit hardware resources. Figure~\ref{fig:unroll} demonstrates that our manual optimization of loop unrolling and instruction scheduling outperforms the compiler's automatic handling by $1.1\times$. 

\begin{figure}[t]
\centerline{\includegraphics[width=0.4\textwidth]{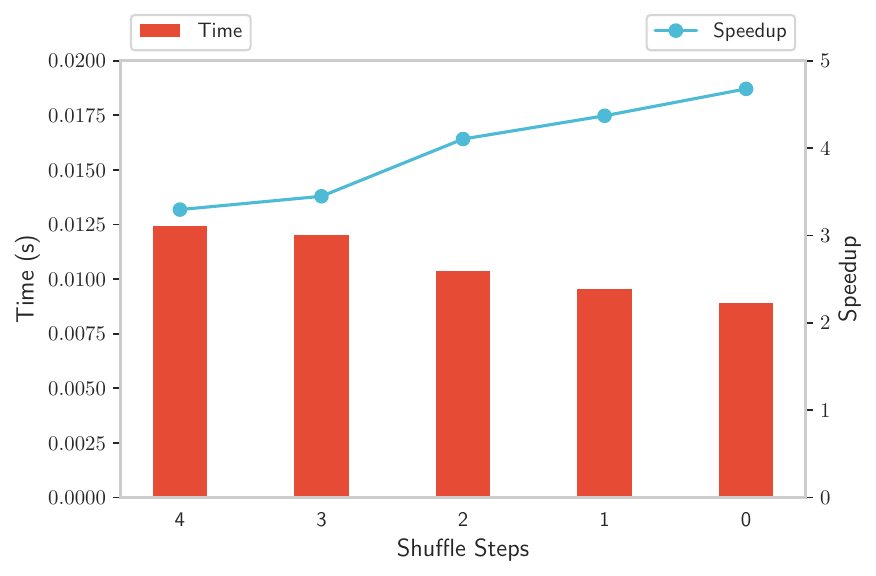}}
\caption{Time consumed and speedup comparing with GCC auto-vectorization version code as shuffle steps go down.}
\label{fig:shuf}
\end{figure}

\subsubsection{Shuffle Steps}

For a fixed tensor shape, we construct scenarios with different shuffle steps by altering the permutation map to test performance. Figure~\ref{fig:shuf} shows that the time required for transposition decreases as the number of shuffle steps reduces. Compared to the case with 4 shuffle steps, the time only reduces by 25\% in the scenario with no shuffle. This indicates that the load-shuffle-store structure within each loop body can mask a large portion of the pipeline time associated with shuffle operations. Additionally, shuffle operations can help fill the pipeline bubbles of the vload instructions. Our approach effectively hides pipeline stalls during data loading, thereby improving overall performance. Moreover, the results support our prediction that the impact of multiple shuffle steps on the runtime will not grow linearly. The additional $log_2w$ shuffle steps do not mean that the time will increase by $log_2w$ times. 

\begin{figure}[t]
\centerline{\includegraphics[width=0.5\textwidth]{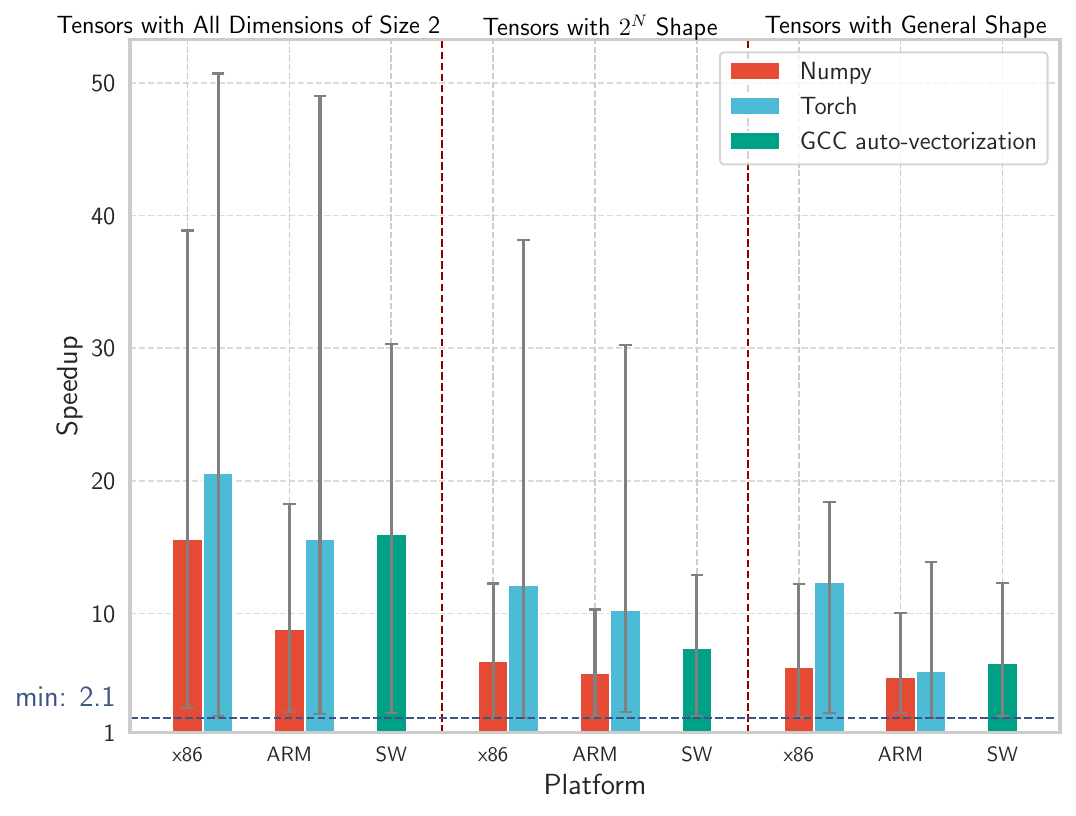}}
\caption{The performance results of cross-platform testing. The left column represents scenario of tensors with all dimensions of size 2, the middle column represents scenario of tensors with $2^N$ shape, and the right column represents scenario of tensors with general shape. Comparisons are made with NumPy and PyTorch on x86 and ARM platforms, and with the GCC auto-vectorized version on Sunway platform. Tensors with moderate size are chosen. The top of the error bar represents the maximum value in the test, while the bottom represents the minimum value.}
\label{fig:cross}
\end{figure}

\subsubsection{Cross-Platform Performance}

We conducted performance tests of the code generator on x86, ARM, and Sunway platforms. Figure~\ref{fig:cross} illustrates the results of cross-platform testing with situations of tensors with all dimensions of size 2, with $2^N$ shape, and with general shape separately. Although there are slight variations in performance across different platforms for different tensor shapes, the overall trend remains consistent. In some specific cases, an acceleration factor of up to $50 \times$ relative to \texttt{NumPy} can be achieved. The minimum acceleration factor for different tensor shapes tested across various platforms is $2.1 \times$, which demonstrates the robust performance stability of our cross-platform code generator. For sw-python is not compatible in slave core, we use the GCC auto-vectorization code as baseline on Sunway platform. The results indicate that our generated code achieves good performance across different architectures and different tensor shapes, demonstrating its portability and adaptability.

\section{Conclusion}

In this paper, we have presented \system, a novel SIMD permutation toolkit that addresses the longstanding challenges of efficient tensor permutation across diverse computational domains. By leveraging Single Instruction Multiple Data (SIMD) architectures, \system achieves deterministic performance with a worst-case time complexity of $O(\frac{N \log_2 w}{w})$, offering a significant improvement over previous methods that often lack complexity guarantees or are limited to specific tensor shapes and permutation patterns. Our toolkit dynamically adapts to varying instruction sets, bit widths, tensor configurations, and data types, ensuring high vector utilization and pipeline-level optimization without restrictive assumptions. Compared to established frameworks like HPTT\cite{springer2017hptt}, NumPy, and PyTorch, \system demonstrates remarkable speedups of up to $38\times$ in specialized cases and $5\times$ in general scenarios. 

This work successfully explores the problem of spatial locality preservation in complex tensor permutations. It not only overcomes the limitations of prior vectorized permutation approaches, but also establishes a scalable, high-performance solution for tensor permutations, paving the way for enhanced efficiency in neural networks, tensor networks, and other data-intensive applications. As computational workloads continue to evolve, \system stands as a versatile and robust foundation for future advancements in tensor permutation.


\bibliographystyle{ACM-Reference-Format}
\bibliography{main}

\appendix

\end{document}